\documentclass[preprint]{jpsj2}
\usepackage{graphicx}

\newcommand{\hG}{\hat{G}}
\newcommand{\ene}{\varepsilon}
\newcommand{\vq}{{\mib q}}
\newcommand{\vr}{{\mib r}}
\newcommand{\LL}{L}
\newcommand{\RL}{R}

\def\runtitle{Effects of Polaron Formation in Quantum Dots}
\def\runauthor{T.\ {\sc Tasai} and M.\ {\sc Eto}}

\title{%
Effects of Polaron Formation in Semiconductor Quantum Dots on \\
Transport Properties
}

\author{%
Tomoki {\sc Tasai} and Mikio {\sc Eto}\thanks{Corresponding author.
E-mail address: eto@rk.phys.keio.ac.jp}
}

\inst{%
Faculty of Science and Technology, Keio University, \\
3-14-1 Hiyoshi, Kohoku-ku, Yokohama 223-8522, Japan
}

\recdate{January 23, 2003}

\abst{%
We theoretically examine the effects of polaron formation in quantum
dots on the transport properties. When a separation between two electron-levels
in a quantum dot matches the energy of the longitudinal optical (LO) phonons,
the polarons are strongly formed. The Rabi splitting between the levels
is observable in a peak structure of the differential conductance $G$
as a function of the bias voltage. The polaron formation
suppresses the peak height of $G$, which is due to the
competition between the resonant tunneling (resonance between a
level in the dot and states in the leads) and the polaron formation
(Rabi oscillation between two levels in the dot).
$G$ shows a sharp dip at the midpoint between the split peaks.
This is attributable to the destructive interference between
bonding and anti-bonding states in a composite system of electrons
and phonons.
}

\kword{%
quantum dot, polaron, resonant tunneling, LO phonon, Rabi splitting,
self-consistent Born approximation
}

\begin{document}
\sloppy
\maketitle

\section{Introduction}
\label{sec1}

In semiconductor quantum dots, the relaxation of electrons between
discrete energy-levels is an important issue from a viewpoint of
application to quantum-dot lasers.\cite{Arakawa}
The electron-phonon interaction should be much less efficient for the
relaxation than in the bulk. This is because the number of phonons is
very limited whose energies match the separation between the 
discrete levels,\cite{Bastard0,Benisty}
which is in contrast to the case of bulk where the electron levels form a
continuous spectrum. Related to this ``phonon bottleneck,'' which is not
always observed in actual quantum dots, several theoretical works have been
reported on the relaxation mechanisms in quantum
dots.\cite{relax0,relax1,relax2,relax3,relax4,relax5}

A remarkable exception of the inefficiency of the electron-phonon
interaction is seen when the energy of the longitudinal optical
(LO) phonons matches the level separation.
Then the interaction can be significantly strong, reflecting a
large density of states of the LO phonons with little dispersion.
The electron-LO-phonon interaction results in the formation of
``polarons.'' The polarons in quantum dots are coherent states
consisting of the discrete electron-levels and LO phonons, which
were predicted by Inoshita and Sakaki.\cite{Inoshita}
The polarons have been found in self-assembled InAs quantum dots
embedded in GaAs, using the far-infrared spectroscopy.\cite{Bastard1,Bastard2}
When a spacing between the electron levels is tuned to match the
energy of LO phonons by applying magnetic fields, the Rabi splitting
of the coherent states has been observed in the absorption spectra.

In the present paper, we theoretically examine a situation to observe the
polarons by the electric transport, considering a quantum dot connected to
external leads. The conductance takes place by the resonant tunneling
through electron levels in the quantum dot. We show that the polaron formation
is observable in a peak structure of the differential conductance $G$
as a function of the bias voltage. Recently the transport properties of
one of the self-assembled InAs quantum dots can be investigated in
vertically fabricated samples of
heterostructures including a layer of such quantum dots.\cite{QD1,QD2}
Hence our theoretical results are expected to be observed experimentally.
Compared with the optical experiments, the transport measurement has
the advantage to be able to examine the polaron formation in a single
quantum dot. Hence the measurement could reveal its characters
in more details.

From a viewpoint of the fundamental research, this is a simple example
of important problems for the coupling between electrons and bosonic fields
in the environment. We find new interference effects on the conductance
in a composite system of electrons and phonons. (i) The polaron formation
suppresses the peak height of $G$, which is due to the
competition between the resonant tunneling (resonance between a
level in the dot and states in the leads) and the polaron formation
(Rabi oscillation between two levels in the dot). (ii)
$G$ shows a sharp dip at the midpoint between the split peaks.
This is attributable to the destructive interference between
bonding and anti-bonding states in the composite system.
This is analogous to the transport through double quantum dots
connected in parallel\cite{Kawamura} and Fano resonance.\cite{Fano}

The organization of the present paper is as follows. In section 2,
we explain our models and calculation methods.
In section 3, we examine an exactly solvable model. By the calculations
using Green's
function, we show that the polaron formation is observable in a
peak structure of the differential conductance.
In section 4, we consider a more general model.
The Green's function is calculated in the self-consistent Born
approximation to take into account the electron-phonon interaction
to infinite orders. The new interference effects on the conductance are
discussed. The last section (section 5) is devoted to the
conclusions and discussion.

\section{Models and Calculation Methods}
\label{sec2}

We consider a quantum dot with two levels, $g$ and $e$,
for electrons, as shown in Fig.\ 1(a). The level spacing,
$\Delta\varepsilon=\varepsilon_e-\varepsilon_g$, is tuned.
The levels are coupled to two external leads, $\LL$, $\RL$,
through tunnel barriers. The Hamiltonian of electrons,
$H_{\rm e}+H_{\rm T}$, is given by
\begin{eqnarray}
H_{\rm e} & = & \sum_{i=g,e} \varepsilon_i d_i^{\dagger}d_i
   +\sum_{\alpha=\LL,\RL}\sum_{k}
    \varepsilon_k a_{\alpha,k}^{\dagger}a_{\alpha,k}, \\
H_{\rm T} & = & \sum_{i=g,e}\sum_{\alpha=\LL,\RL}\sum_{k}
               (V^i_{\alpha}d_i^{\dagger}a_{\alpha,k}+{\rm h.c.}),
\end{eqnarray}
where $d_i^{\dagger}$, $a_{\alpha,k}^{\dagger}$ ($d_i$, $a_{\alpha,k}$)
are the creation (annihilation) operators for an electron in the dot
and in lead $\alpha$, respectively. The spins of electrons are disregarded
since they are not relevant in this problem. The electron-electron
interaction is not taken into account in this paper. The tunnel couplings,
$H_{\rm T}$, broaden
the level $i$ in the dot with $\Delta^i=\Delta^i_{\LL}+\Delta^i_{\RL}$,
\begin{equation}
\Delta^i_{\alpha}=\pi \nu |V^i_{\alpha}|^2,
\end{equation}
where $\nu$ is the density of states in the leads.\cite{com1}

For phonons, we consider an LO phonon mode,
\begin{equation}
H_{\rm ph} = \sum_{\vq}
          \hbar\omega_{\vq} b_{\vq}^{\dagger}b_{\vq},
\end{equation}
where $b_{\vq}^{\dagger}$ ($b_{\vq}$) creates (annihilates) an LO phonon
with momentum $\vq$.
Since only phonons with long wavelength, $|{\mib q}|_{\sim}^< 2\pi/$(dot size),
interact with electrons significantly,\cite{Inoshita}
we can regard the phonon mode as dispersionless;
$\omega_{\mib q}=\omega_{\rm LO}$.
Other modes of phonons are neglected since they do not effectively couple
to electrons in the quantum dot when $\Delta\ene$ is of the order
of $10$meV\cite{Bastard1,Bastard2} (Appendix A).

The electron-LO-phonon interaction is described by
the Fr\"ohlich Hamiltonian $H_{\rm e-ph}$ (Appendix A).
For the phonon-absorption process, the matrix element between an electron
at level $e$ with  $N-1$ phonon, $|e, (N-1)_{\mib q} \rangle$, and an
electron at $g$ with $N$ phonon, $|g, N_{\mib q} \rangle$, is written as
\begin{equation}
\langle e, (N-1)_{\mib q} | H_{\rm e-ph} | g, N_{\mib q} \rangle
= \sqrt{N_{\mib q}} v_{\mib q}.
\label{eq:eph}
\end{equation}
We denote
\begin{equation}
v^2=\sum_{\vq} |v_{\vq}|^2
\end{equation}
which characterizes the strength of the electron-phonon interaction.
We set $v=0.2\hbar\omega_{\rm LO}$, considering the experimental
situations.\cite{Bastard1,Bastard2}

The total Hamiltonian is
\begin{equation}
H_{\rm tot}=H_{\rm e}+H_{\rm T}+H_{\rm ph}+H_{\rm e-ph}.
\end{equation}
To examine the polaron formation, we have to include $H_{\rm e-ph}$ to
all orders. For the purpose, (i) we study an
exactly solvable model in which only level $e$ in the dot couples to the
leads by the tunneling; $V^g_{\LL}=V^g_{\RL}=0$.
We present the Green's function in an analytical form in section 3.
(ii) In general situations, we take $H_{\rm e}+H_{\rm T}+H_{\rm ph}$
as an unperturbed Hamiltonian,
taking account of the tunnel couplings between the dot and
leads exactly. Then we consider $H_{\rm e-ph}$ as a perturbation, using the
self-consistent Born approximation\cite{Mahan} (section 4).

\section{Calculated Results with Solvable Model}

To illustrate the effects of polaron formation on the transport properties,
we begin with a simple model with $V^g_{\LL}=V^g_{\RL}=0$.
This model is exactly solvable if we restrict the states in the quantum
dot to $|e, 0_{\rm ph} \rangle$ (an electron at level $e$ and
no phonon), $|g, 1_{\mib q} \rangle$ (an electron at level $g$ and
one phonon), and no electron nor phonon (Fig.\ 1(b)).

The Green's function $\hG(\varepsilon)$, which is defined by
\begin{equation}
 (\varepsilon -H_{\rm tot}   + i \delta ) \hG(\varepsilon)  = 1,
\label{eq:Green}
\end{equation}
is obtained in an analytical way, as shown in Appendix B.
The diagonal elements for the states in the dot are
\begin{eqnarray}
G_{(e, 0_{\rm ph})} (\ene) & = & 
\langle e, 0_{\rm ph} | \hG(\varepsilon) | e, 0_{\rm ph} \rangle \nonumber \\
& = & \frac{1}{\varepsilon-\varepsilon_e
               +i\Delta^e-\Sigma_{(e, 0_{\rm ph})}(\ene)},
\label{eq:Green1} \\
\Sigma_{(e, 0_{\rm ph})}(\ene)
& = & \frac{v^2}
{\varepsilon-(\varepsilon_g+\hbar\omega_{\rm LO})+i\delta},
\label{eq:selfE1}
\end{eqnarray}
and
\begin{eqnarray}
G_{(g, 1_{\mib q})}(\ene) & = & 
\langle g, 1_{\mib q} | \hG(\varepsilon) | g, 1_{\mib q} \rangle \nonumber \\
& = & \frac{1}{\varepsilon-(\varepsilon_g+\hbar\omega_{\rm LO})
              -\Sigma_{(g, 1_{\mib q})}(\ene)},
\label{eq:Green2} \\
\Sigma_{(g, 1_{\mib q})}(\ene)
& = &  \frac{v^2}
{\varepsilon-\varepsilon_e+i\Delta^e-
 \frac{v^2}{\varepsilon-(\varepsilon_g+\hbar\omega_{\rm LO})+i\delta}}.
\label{eq:selfE2}
\end{eqnarray}

In terms of the Green's function, the density of states in the dot
is written as
\begin{equation}
D(\varepsilon) = -\frac{1}{\pi}{\rm Im} \left[
         G_{(e, 0_{\rm ph})}(\ene)
        +\sum_{\mib q}G_{(g, 1_{\mib q})}(\ene) \right].
\end{equation}
The electric current from lead $L$ to $R$ is
\begin{eqnarray}
I=e\frac{2\pi}{\hbar}\sum_{k,k'}
|V^{e *}_{\RL}G_{(e, 0_{\rm ph})}(\varepsilon_k) V^e_{\LL}|^2
\delta(\varepsilon_{k'}-\varepsilon_k) \nonumber \\
\hspace*{2.6cm}
\times [f(\varepsilon_k-\mu_{\LL})-f(\varepsilon_{k'}-\mu_{\RL})],
\label{eq:current}
\end{eqnarray}
where $\mu_{\alpha}$ is the Fermi level in lead $\alpha$ and
$f(\varepsilon)$ is the Fermi distribution function. At $T=0$,
the differential conductance is given by
\begin{eqnarray}
G & \equiv & e\frac{dI}{d\mu_{\LL}} \nonumber \\
  & = & \frac{e^2}{h}
        \left.
        \frac{4\Delta^e_{\LL}\Delta^e_{\RL}}
       {(\ene-\varepsilon_e -{\rm Re}\Sigma_{(e, 0_{\rm ph})}(\ene))^2
        +(\Delta^e-{\rm Im}\Sigma_{(e, 0_{\rm ph})}(\ene))^2}
        \right|_{\ene=\mu_{\LL}},
\label{eq:conductance}
\end{eqnarray}
when $\mu_{\RL}$ is fixed.

In Fig.\ 2, we show the calculated results of (a) the density of states
in the quantum dot and (b) differential conductance $G$ at $T=0$.
The level spacing,
$\Delta\ene=\varepsilon_e-\varepsilon_g$, is changed gradually.
In (a), two Lorentzian peaks appear corresponding to the
states $|g, 1_{\mib q} \rangle$ and $|e, 0_{\rm ph} \rangle$. The peak
of the former ($\ene \approx \ene_g+\hbar\omega_{\rm LO}$) is
sharp and high, reflecting large density of states
of the LO phonons. The peak of the latter ($\ene \approx \ene_g+\Delta\ene$)
is more broadened owing to the tunnel coupling to the leads.
Around $\Delta\varepsilon=\hbar\omega_{\rm LO}$,
an anti-crossing between the states is seen,
which is ascribed to the formation of polarons.
The electron-phonon interaction mixes the states
$|g, 1_{\mib q} \rangle$ and $|e, 0_{\rm ph} \rangle$, making the bonding
state with lower energy and the anti-bonding state with higher energy.
Hence this anti-crossing of the two peaks is due to the Rabi splitting caused
by the electron-phonon interaction.

The differential conductance $G$ shows a similar peak structure,
as a function of the bias voltage, $(\mu_{\LL}-\mu_{\RL})/e$,
to that in the density of states $D(\ene)$.
The current flows by the resonant tunneling
through the electronic states in the dot. Because of the mixture between
$|g, 1_{\mib q} \rangle$ and $|e, 0_{\rm ph} \rangle$, $G$ has a peak even
at the position of states $|g, 1_{\mib q} \rangle$ in spite of the absence
of the tunnel coupling of level $g$. In conclusion, we clearly observe the
Rabi-splitting by the polaron formation in the transport properties of
the quantum dots.

\section{Calculated Results in Self-Consistent Born Approximation}

When both the levels in the quantum dot couple to the leads, we
adopt the self-consistent Born approximation
to consider the electron-phonon interaction to infinite orders
(Appendix C). In this approximation,\cite{Mahan} the self-energy
$\Sigma_{i}(\varepsilon)$ in Green's function of
electrons,
\begin{equation}
G_i(\ene)=\frac{1}{\ene-\ene_i+i\Delta^i-\Sigma_{i}(\ene)}
\label{eq:SCB1}
\end{equation}
$(i=g,e)$, is determined by the self-consistent equation
\begin{eqnarray}
\Sigma_i(\ene) = \sum_{\vq} |v_{\mib q}| ^2
     \left[ \frac{N_{\mib q}}{\ene + \hbar\omega_\vq- \ene_j +i\Delta^j
                  -\Sigma_j(\ene + \hbar\omega_\vq)} \right. \nonumber \\
     \left. +\frac{N_\vq+1}{\ene - \hbar\omega_\vq - \ene_j +i\Delta^j
                  -\Sigma_j(\ene - \hbar\omega_\vq)} \right].
\label{eq:SCB2}
\end{eqnarray}
Here, $j \ne i$ and $N_{\mib q}$ is the number of phonons with momentum
${\mib q}$.\cite{com2}
Note that we disregard the inelastic transport processes where
an electron enters a level in the dot and goes out from
the other level in the dot,
to focus on the effects of the polaron formation on the transport.
This is a good approximation when $\Delta^g_{L,R} \ll \Delta^e_{L,R}$.
We will discuss this approximation later.
The density of states in the dot, $D(\ene)$, and the differential
conductance, $G$, are calculated in the same way as in the previous
section.

Figure 3 presents the calculated results at $T=0$. The density of states in
the dot shows three large peaks at $\ene \approx \ene_g$,
$\ene_g+\hbar\omega_{\rm LO}$, $\ene_g+\Delta\ene$, which correspond to
`level $g$ with no phonon,' `level $g$ with one phonon' and
`level $e$ with no phonon,' respectively. A small peak at
$\ene \approx \ene_g+\Delta\ene+\hbar\omega_{\rm LO}$ is also observable,
corresponding to `level $e$ with one phonon.' The peak widths are determined
mainly by the level broadening, $\Delta^{g,e}$, due to the tunnel couplings
to the leads. An anti-crossing between $|g, 1_{\vq} \rangle$ and
$|e, 0_{\rm ph} \rangle$ is seen at $\Delta\ene \approx \hbar\omega_{\rm LO}$,
as discussed in Fig.\ 2.

The differential conductance $G$ through the dot shows a similar
peak structure as a function of the bias voltage. Now we discuss two new
interference effects of the polaron formation on $G$.
First, the polaron formation suppresses the peak height of $G$.
When $|\Delta\ene - \hbar\omega_{\rm LO}|$ is large, the resonant
tunneling through level $g$ or $e$ makes peaks of $G \approx e^2/h$
in height. When $\Delta\ene \approx \hbar\omega_{\rm LO}$, however,
$G$ is much smaller than $e^2/h$. This is because the electron-phonon
interaction results in a finite life-time of the electron levels, $\tau$
(Im$\Sigma_i \sim -\hbar/\tau$),
which weakens the resonant tunneling by a factor of
$\sim [\Delta^{i}/(\Delta^i+\hbar/\tau)]^2$.\cite{com4}

Second, the differential conductance shows a dip at the midpoint
between the two peaks. To see this dip more clearly, we show the
conductance at $T=0$ with larger level broadening $\Delta^e$ in
Fig.\ 4. The level spacing in the dot is fixed to match the energy of
the LO phonons; $\Delta \ene=\hbar\omega_{\rm LO}$.
This dip is sharp and deep when $\Delta^g \ll \Delta^e$.
(In model (a) with $\Delta^g=0$, the analytical solution indicates that
$G=0$ exactly at $\mu_L=\ene_g+\hbar\omega_{\rm LO}$.)
With increasing $\Delta^g/\Delta^e$, the dip is smeared.
This dip of $G$ is a consequence of the destructive
interference between the bonding and anti-bonding states of
$|g, 1_{\vq} \rangle$ and $|e, 0_{\rm ph} \rangle$.
Kawamura and Aono have proposed a similar dip of $G$ in the transport
through double quantum dots coupled in parallel.\cite{Kawamura}
They have considered a situation where only one of the dots is
connected to the leads by the tunneling. In this case,
the differential conductance has two peaks due to the resonant
tunneling through
bonding and anti-bonding orbitals between the dots. $G$ has a dip
at the midpoint between the peaks ($G=0$ exactly when the tunnel
couplings to left and right leads are equivalent).
Another analogous phenomenon is the Fano resonance.\cite{Fano}
The interference between a discrete level and a continuum of states
results in an
asymmetric shape of the resonance, with a dip on one side of the
resonance. This dip is due to the destructive interaction between
bonding and anti-bonding couplings of the discrete level and the
continuum. Note that, in our model, the dip of $G$ is attributable
to the destructive interference in a composite system of
electrons and phonons.

Finally, we discuss the differential conductance $G$ at finite temperature.
When $k_{\rm B}T \gg \Delta^{g,e}$, the peak heights of $G$ are much smaller
than $e^2/h$, whereas the peak widths are determined by the thermal energy.
Figure 5 shows $G$ at
$k_{\rm B}T=0.05 \hbar\omega_{\rm LO}$ ($T \approx 18$K),
$\Delta^g_{\LL}=\Delta^g_{\RL}=0.005\hbar\omega_{\rm LO}$ and
$\Delta^e_{\LL}=\Delta^e_{\RL}=0.01\hbar\omega_{\rm LO}$.
Although the peaks of $G$ are thermally broadened,
the evidence of the polaron formation is still observable.
We can also see the above-mentioned characters of $G$, the suppression of
the peak heights at $\Delta\ene \approx \hbar\omega_{\rm LO}$ and
the dip of $G$ between the resonant peaks.

\section{Conclusions and Discussion}

We have investigated the effects of polaron formation in a quantum
dot on the transport properties. When a separation between two levels
in the dot, $g$ and $e$, matches the energy of the LO phonons,
the polarons are strongly formed.
The Rabi splitting between states $|g, 1_{\vq} \rangle$ and
$|e, 0_{\rm ph} \rangle$ is observable in a peak structure of the
differential conductance $G$, as a function of the bias voltage.
We have found new interference effects on the transport in this
composite system of electrons and phonons. The polaron formation
suppresses the peak height of $G$, which is attributable to the
competition between the resonant tunneling (resonance between a
level in the dot and states in the leads) and polaron formation
(Rabi oscillation between two levels in the dot).
$G$ shows a sharp dip at the midpoint between the split peaks.
This is due to the destructive interference between
bonding and anti-bonding states of
$|g, 1_{\vq} \rangle$ and $|e, 0_{\rm ph} \rangle$.
These theoretical results are expected to be observed experimentally.
The measurement of the electric current would enable to observe the polaron
formation in a single quantum dot and reveal its characters
in more details.

In the calculations with the self-consistent Born approximation,
we have disregarded inelastic transport processes: When an electron is
injected to a level in the dot, say $e$, from lead $\LL$, the electron
is always ejected from the same level, $e$, to lead $\RL$. This is a good
approximation when $\Delta^g_{L,R} \ll \Delta^e_{L,R}$. (Indeed the
upper level usually couples to the leads more strongly than the lower
level in quantum dots.) Otherwise, the relaxation from
level $e$ to $g$ could take place in transport processes, which would
broaden the peaks of the differential conductance. To consider the
inelastic processes, we have to develop the self-consistent Born
approximation in the nonequilibrium Green's function method. This
is a challenging problem and requires the further study.
Besides, we have not considered the coupling of electrons to
longitudinal acoustic (LA) phonons\cite{Inoshita} nor
electron-electron interaction
(Auger processes for the relaxation of electrons\cite{relax0}, etc.)
which might cause the decay of polarons and, as a result, broaden
the conductance peaks further. Although the quantitative estimation of
these effects is beyond the scope of the present study,
the polaron formation should
not be influenced by them qualitatively.

This work was partially supported by a Grant-in-Aid for
Scientific Research in Priority Areas ``Semiconductor Nanospintronics''
(No.\ 14076216) of The Ministry of Education, Culture, Sports, Science
and Technology, Japan.

\appendix
\section{Electron-Phonon Interaction in Quantum Dots}

The interaction between electrons and LO phonons is described by the
Fr\"ohlich Hamiltonian.\cite{Mahan} In quantum dots, it is given by
\begin{eqnarray}
H_{\rm e-ph} & = & \frac{1}{\sqrt{V}}
            \sum_{\vq,ij}
            M_{\vq, ij} d_i^{\dagger} d_j
            (b_{\vq}+b_{-\vq}^{\dagger}), \\
M_{\vq, ij} & = &
             i\sqrt{4\alpha_{\rm e-ph}} \frac{\hbar\omega_\vq}{q}
             \left( \frac{\hbar}{2m^* \omega_\vq} \right)^{1/4}
             \langle i | e^{i\vq\cdot\vr} | j \rangle, \\
\alpha_{\rm e-ph} & = & \frac{e^2}{\hbar}
             \sqrt{\frac{m^*}{2\hbar\omega_\vq}}
             \left( \frac{1}{\varepsilon(\infty)} - 
                    \frac{1}{\varepsilon(0)} \right),
\end{eqnarray}
where $| i \rangle$ is the envelope function for electron level $i$
in the dot and $m^*$ is the effective
mass of electrons. $\varepsilon(\infty)$ and $\varepsilon(0)$ are
the dielectric constants at high and low frequencies, respectively.
We assume that phonons are the same as those in the bulk GaAs
($V$ is the volume of the bulk system) in which the InAs quantum
dots are embedded.
The dimensionless coupling constant $\alpha_{\rm e-ph}$ should be
$0.15$ to explain the experimental results,\cite{Bastard1,Bastard2}
whereas $\alpha_{\rm e-ph}=0.06$ in bulk InAs.

Owing to the factor of $\langle i | e^{i\vq\cdot\vr} | j \rangle$,
the electron-phonon interaction is negligibly small when
$|\vq|>2\pi/L$ with $L$ being the dot size
($\sim 20$nm\cite{Bastard1,Bastard2}).
Hence we can regard the LO phonons as dispersionless;
$\hbar\omega_\vq=\hbar\omega_{\rm LO}=36$meV.

In this paper, we consider two electron levels ($g,e$).
The coupling constant for the interlevel interaction is denoted by
$\displaystyle v_\vq=M_{\vq, e,g}/\sqrt{V}$ in the text, Eq.\ (\ref{eq:eph}).
The strength of the couplings is characterized by
\begin{eqnarray}
v^2 & = & \sum_{|\vq| \le 2\pi/L} |v_{\vq}|^2 \nonumber \\
    & = & \frac{1}{V} \sum_{|\vq| \le 2\pi/L} |M_{\vq, e,g}|^2  \nonumber \\
    & = & \frac{4\alpha_{\rm e-ph}}{\pi}(\hbar \omega_{\rm LO})^2
          \sqrt{\frac{\hbar^2}{2m^*L^2}\frac{1}{\hbar \omega_{\rm LO}}}.
\end{eqnarray}
In the above-mentioned situation, $v/(\hbar \omega_{\rm LO}) \approx 0.2$.
We neglect the intralevel interaction, $M_{\vq, g,g}$, $M_{\vq, e,e}$,
assuming that these effects are involved in the self-energies of
electrons.

We disregard the LA phonons for the following reasons.
(i) LA phonons with $|\vq|>2\pi/L$ hardly couple to the electrons
for the same reasons as LO phonons.
(ii) LA phonons with $|\vq| \le 2\pi/L$ are not relevant for the polaron
formation since the energies ($^<_{\sim} 2$meV) are much smaller than
the level spacing of electrons ($\Delta\ene \sim 10$meV).

\section{Calculations of Green's Function (1)}

In the exactly solvable model with $V^g_{\LL}=V^g_{\RL}=0$
(Fig.\ 1(b)),
we divide the total Hamiltonian as
$H_{\rm tot} = H_0+V$, where $H_0 = H_{\rm e}+H_{\rm ph}$ and
$V=H_{\rm T}+H_{\rm e-ph}$. Using Eq.\ (\ref{eq:Green}),
\begin{equation}
\langle e, 0_{\rm ph} | (\varepsilon-H_0+i\delta) \hG | e, 0_{\rm ph} \rangle
= 1+\langle e, 0_{\rm ph} | V \hG | e, 0_{\rm ph} \rangle,
\end{equation}
which yields
\begin{equation}
(\varepsilon-\varepsilon_e+i\delta)G_{(e, 0_{\rm ph})}
=1+\sum_{\vq} v_{\vq}
   \langle g, 1_{\vq} | \hG | e, 0_{\rm ph} \rangle
   +\sum_{\alpha=\LL,\RL}\sum_k V^e_{\alpha}
   \langle \alpha k, 0_{\vq} | \hG | e, 0_{\rm ph} \rangle.
\label{eq:G1}
\end{equation}
Similarly, using Eq.\ (\ref{eq:Green}), we obtain
\begin{eqnarray}
\left[ \varepsilon-(\varepsilon_g+\hbar\omega_{\vq})+i\delta \right]
\langle g, 1_{\vq} | \hG | e, 0_{\rm ph} \rangle
& = & v_{\vq}^* G_{(e, 0_{\rm ph})} 
\label{eq:G2} \\
(\varepsilon-\varepsilon_k+i\delta)
\langle \alpha k, 0_{\vq} | \hG | e, 0_{\rm ph} \rangle
& = & V^{e *}_{\alpha} G_{(e, 0_{\rm ph})}.
\label{eq:G3}
\end{eqnarray}
Equations (\ref{eq:G1}), (\ref{eq:G2}) and (\ref{eq:G3}) yield
\begin{equation}
G_{(e, 0_{\rm ph})} (\ene) = 
  \frac{1}{\varepsilon-\varepsilon_e
           -\Sigma_{\rm T}(\ene)-\Sigma_{(e, 0_{\rm ph})}(\ene)},
\end{equation}
where
\begin{eqnarray}
\Sigma_{\rm T}(\ene) & = & \sum_{\alpha=\LL,\RL}\sum_k
\frac{|V^e_{\alpha}|^2}{\varepsilon-\varepsilon_k+i\delta}
\nonumber \\
 & = & \sum_{\alpha=\LL,\RL} |V^e_{\alpha}|^2 \int \nu d\varepsilon_k
 \left[ P\frac{1}{\varepsilon-\varepsilon_k}-
      i\pi\delta(\varepsilon-\varepsilon_k)  \right]
\nonumber \\
 & = & -i(\Delta^e_{\LL}+\Delta^e_{\RL}),
\end{eqnarray}
and
\begin{eqnarray}
\Sigma_{(e, 0_{\rm ph})}(\ene) & = & \sum_{\mib q} \frac{|v_{\mib q}|^2}
{\varepsilon-(\varepsilon_g+\hbar\omega_{\rm q})+i\delta} \nonumber \\
& = & \frac{v^2}
{\varepsilon-(\varepsilon_g+\hbar\omega_{\rm LO})+i\delta}
\end{eqnarray}
(Eqs.\ (\ref{eq:Green1}) and (\ref{eq:selfE1}) in the text).
The real part of $\Sigma_{\rm T}(\ene)$ can be included in the renormalization
of the energy level $\varepsilon_e$, and hence it is omitted here.

In the same way, we obtain $G_{(g, 1_{\mib q})}(\ene)$ in an analytical
form of Eqs.\ (\ref{eq:Green2}) and (\ref{eq:selfE2}).
The electric current is written as Eq.\ (\ref{eq:current}) because
\begin{equation}
\langle {\RL} k', 0_{\rm ph} | \hat{T}(\ene) | {\LL} k, 0_{\rm ph} \rangle
=V^{e *}_{\RL}G_{(e, 0_{\rm ph})}(\varepsilon) V^e_{\LL},
\end{equation}
where the T-matrix is given by
$\hat{T}(\ene)=H_{\rm T}+H_{\rm T}\hG (\ene) H_{\rm T}$.

\section{Calculations of Green's Function (2)}

When both the levels in the quantum dot couple to the leads, the
Green's function cannot be obtained analytically. Then we divide the
Hamiltonian as $H_{\rm tot} = H_0+H_{\rm e-ph}$ with
$H_0 = H_{\rm e}+H_{\rm ph}+H_{\rm T}$. First, we calculate
the unperturbed Green's function,
$(\varepsilon -H_0 + i \delta ) \hG^0(\varepsilon)  = 1$, in the same
way as in Appendix B;
\begin{equation}
G_{i}^0 (\ene) = 
\frac{1}{\varepsilon-\varepsilon_i+i\Delta^i}
\end{equation}
($i=g,e$).\cite{com1} Next, the electron-phonon interaction, $H_{\rm e-ph}$,
is taken into account to infinite orders, by the self-consistent
Born approximation.\cite{Mahan} The calculation yields Eqs.\
(\ref{eq:SCB1}) and (\ref{eq:SCB2}).

At $T=0$, $N_{\mib q}=0$ in Eq.\ (\ref{eq:SCB2}).
$\Sigma_i(\ene)$ is expressed by $\Sigma_i(\ene-\hbar\omega_{\rm LO})$,
which is expressed in turn by
$\Sigma_i(\ene-2\hbar\omega_{\rm LO})$, and so on. These yield
the expression of $\Sigma_i(\ene)$ as a continued fraction.
We find that the continued fraction converges with the truncation
at $\Sigma_i(\ene- 6\hbar\omega_{\rm LO})$.
At $T \ne 0$, $\Sigma_i(\ene)$ is coupled to
$\Sigma_i(\ene\pm \hbar\omega_{\rm LO})$ in Eq.\ (\ref{eq:selfE2}),
$\Sigma_i(\ene\pm \hbar\omega_{\rm LO})$ are coupled to
$\Sigma_i(\ene\pm 2\hbar\omega_{\rm LO})$ and $\Sigma_i(\ene)$, and so on.
These are simultaneous nonlinear equations. We solve the equations
self-consistently in the approximation of
$\Sigma_i(\ene \pm 6\hbar\omega_{\rm LO})=0$ to obtain
$\Sigma_i(\ene)$.

After the calculations of the Green's functions, we evaluate the
density of states of electrons by\cite{com2}
\begin{equation}
D(\varepsilon) = -\frac{1}{\pi}{\rm Im} \left[
         G_{g}(\ene)+G_{e}(\ene) \right].
\end{equation}
The electric current is expressed as
\begin{equation}
I=e\frac{2\pi}{\hbar}\sum_{i=g,e}\sum_{k,k'}
|V^{i *}_{\RL}G_i(\ene_k) V^i_{\LL}|^2
\delta(\varepsilon_{k'}-\varepsilon_k)
[f(\varepsilon_k-\mu_{\LL})-f(\varepsilon_{k'}-\mu_{\RL})].
\end{equation}
The differential conductance is obtained by $G=e dI/d\mu_{\LL}$.
As mentioned in the text, we do not take account of
inelastic transport processes where an electron enters a
level in the dot and goes out from the other level in the dot.

\pagebreak

\vspace*{2cm}

\begin{figure}[hbt]
\begin{center}
\includegraphics[width=10cm]{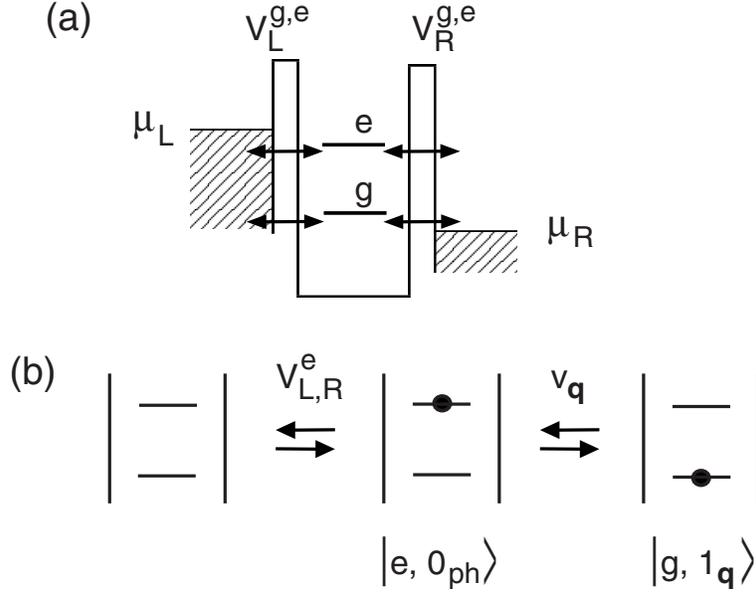}
\end{center}
\caption{(a) We consider a quantum dot with two levels,
$g$ and $e$, in the presence of electron-LO-phonon interaction.
The level spacing, $\Delta\varepsilon=\varepsilon_e-\varepsilon_g$, is tuned.
The levels are coupled to two external leads ($\alpha=$ $L$, $R$) through
tunnel barriers with $V^g_{\alpha}$ and $V^e_{\alpha}$.
(b) In a solvable model with
$V^g_{\LL}=V^g_{\RL}=0$, the states in the dot are restricted to
$|e, 0_{\rm ph} \rangle$ (an electron at level $e$ and
no phonon; middle panel), $|g, 1_{\mib q} \rangle$ (an electron at level $g$
and one phonon; right panel), or no electron nor phonon (left panel).}
\label{fig1}
\end{figure}

\vspace{3cm}

\noindent
{\large Tasai and Eto (Fig.\ 1)}

\pagebreak

\vspace*{2cm}

\begin{figure}[hbt]
\begin{center}
\includegraphics[width=10cm]{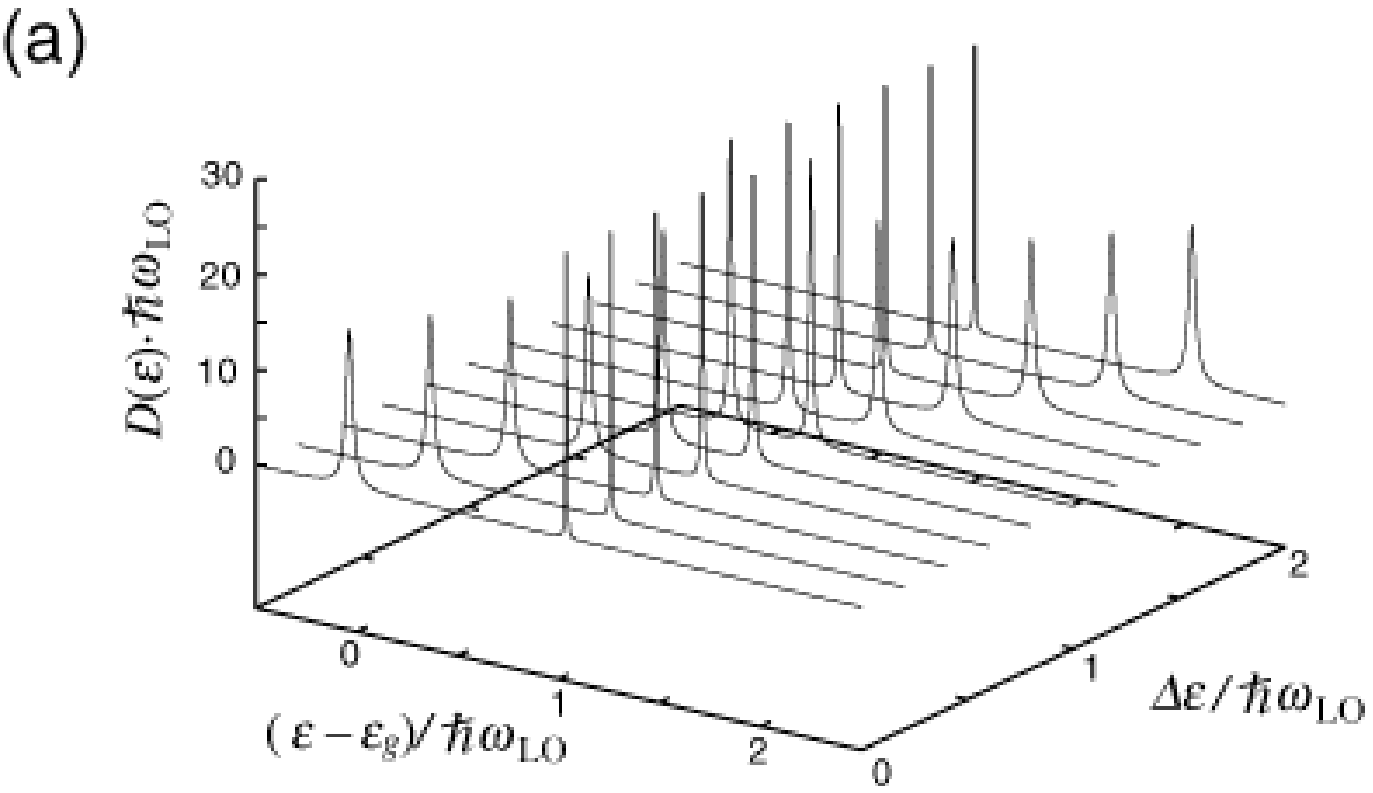}  \vspace{0.5cm} \\
\includegraphics[width=10cm]{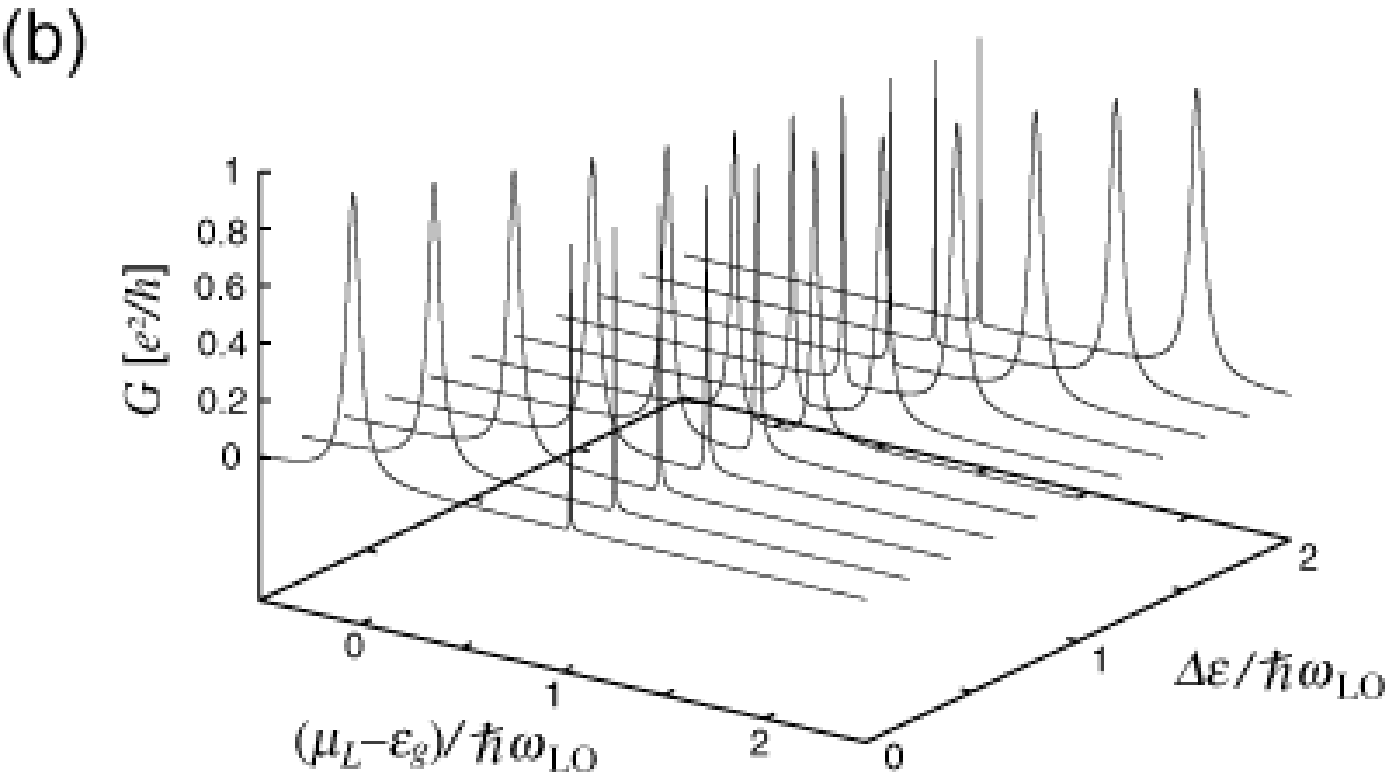}
\end{center}
\caption{The calculated results of an exactly solvable model with
$V^g_{\LL}=V^g_{\RL}=0$. The broadening of level $e$ by the
tunnel coupling to the leads is $\Delta^e_{\LL}=\Delta^e_{\RL}=0.01
\hbar\omega_{\rm LO}$. (a) The density of states in the quantum dot,
$D(\varepsilon)$, with changing the level spacing
$\Delta\ene=\varepsilon_e-\varepsilon_g$.
(b) The differential conductance $G$ through the dot at $T=0$,
as a function of the Fermi level in lead $\LL$, $\mu_{\LL}$,
whereas $\mu_{\RL}$ is fixed.}
\label{fig2}
\end{figure}

\vspace{3cm}

\noindent
{\large Tasai and Eto (Fig.\ 2)}

\pagebreak

\vspace*{2cm}

\begin{figure}[h]
\begin{center}
\includegraphics[width=10cm]{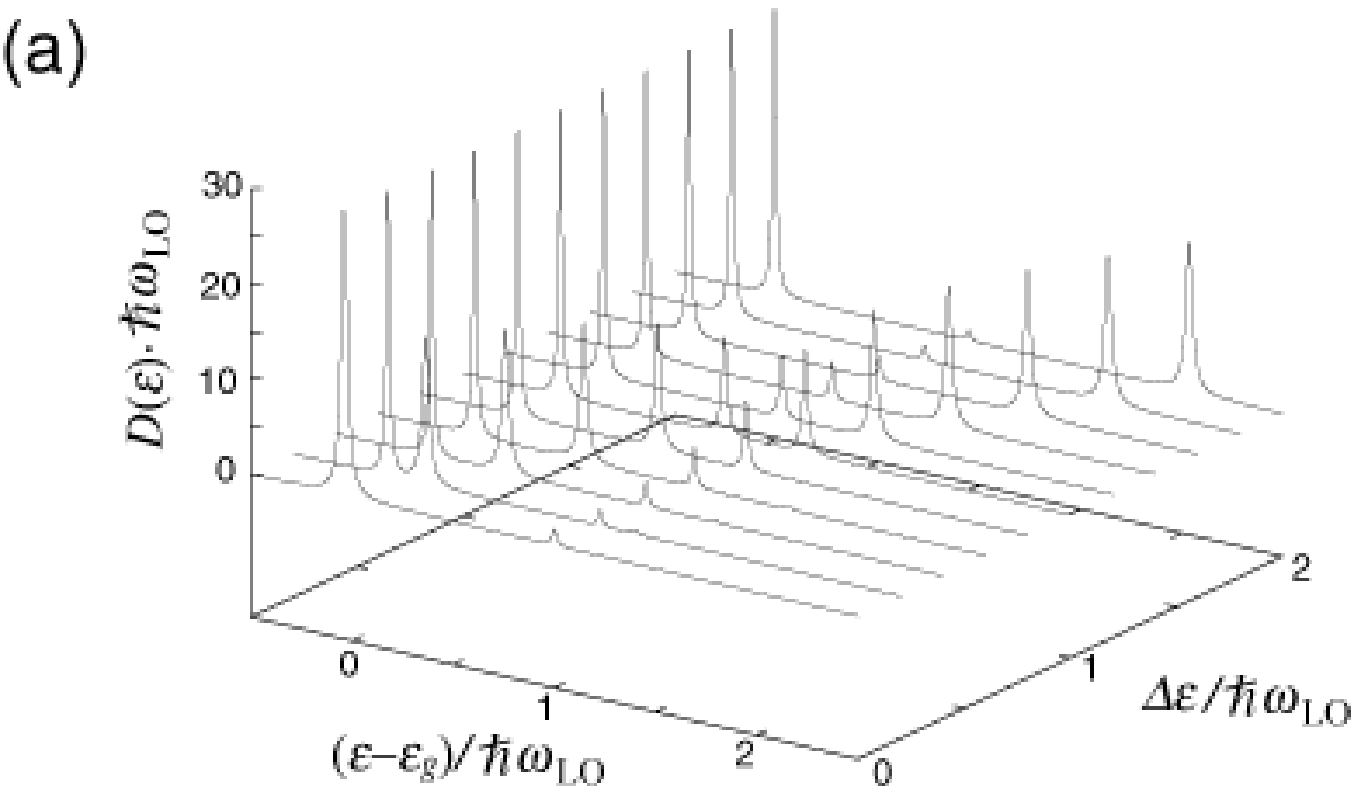} \vspace{0.5cm} \\
\includegraphics[width=10cm]{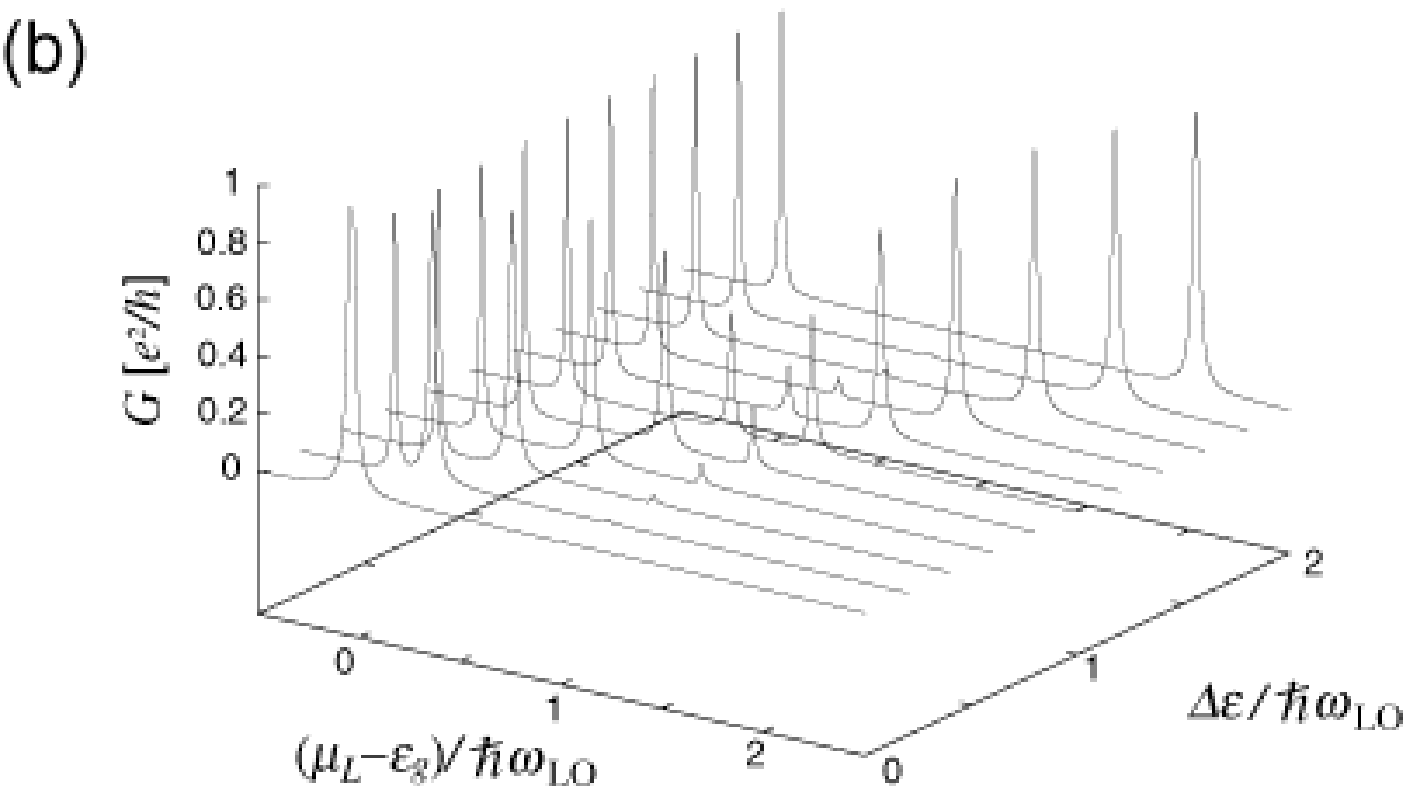}
\end{center}
\caption{The calculated results using the self-consistent Born
approximation. The broadenings of levels $g$ and $e$ by the
tunnel coupling to the leads are
$\Delta^g_{\LL}=\Delta^g_{\RL}=0.005\hbar\omega_{\rm LO}$ and
$\Delta^e_{\LL}=\Delta^e_{\RL}=0.01\hbar\omega_{\rm LO}$,
respectively.
(a) The density of states in the quantum dot,
$D(\varepsilon)$, with changing the level spacing
$\Delta \ene=\varepsilon_e-\varepsilon_g$.
(b) The differential conductance $G$ through the dot at $T=0$,
as a function of the Fermi level in lead $\LL$, $\mu_{\LL}$,
whereas $\mu_{\RL}$ is fixed.}
\label{fig3}
\end{figure}

\vspace{3cm}

\noindent
{\large Tasai and Eto (Fig.\ 3)}

\pagebreak

\vspace*{2cm}

\begin{figure}[hbt]
\begin{center}
\includegraphics[width=8cm]{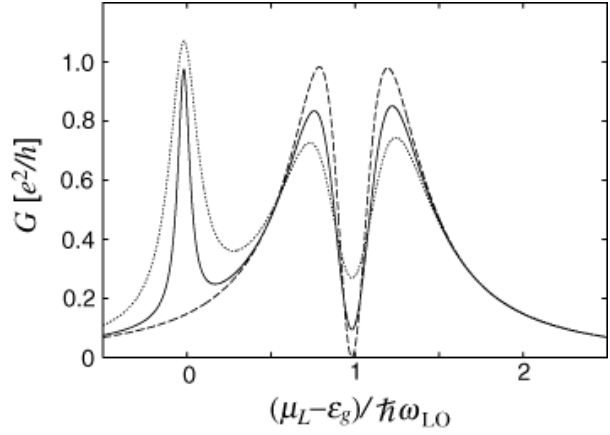}
\end{center}
\caption{The differential conductance $G$ through the quantum dot
at $T=0$, as a function of the Fermi level in lead $\LL$, $\mu_{\LL}$
($\mu_{\RL}$ is fixed).
The level spacing in the dot matches the energy of the LO
phonons; $\Delta \ene=\hbar\omega_{\rm LO}$.
The broadening of the upper level $e$ is fixed at
$\Delta^e_{\LL}=\Delta^e_{\RL}=0.2\hbar\omega_{\rm LO}$,
whereas that of the lower level $g$ is
$\Delta^g_{\LL}=\Delta^g_{\RL}=0.02\hbar\omega_{\rm LO}$ (solid line) and
$0.05\hbar\omega_{\rm LO}$ (dotted line).
The case without the tunnel coupling of
the lower level ($V^g_{\LL}=V^g_{\RL}=0$) is shown by broken line.
}
\label{fig4}
\end{figure}

\vspace{3cm}

\noindent
{\large Tasai and Eto (Fig.\ 4)}

\pagebreak

\vspace*{2cm}

\begin{figure}[h]
\begin{center}
\includegraphics[width=10cm]{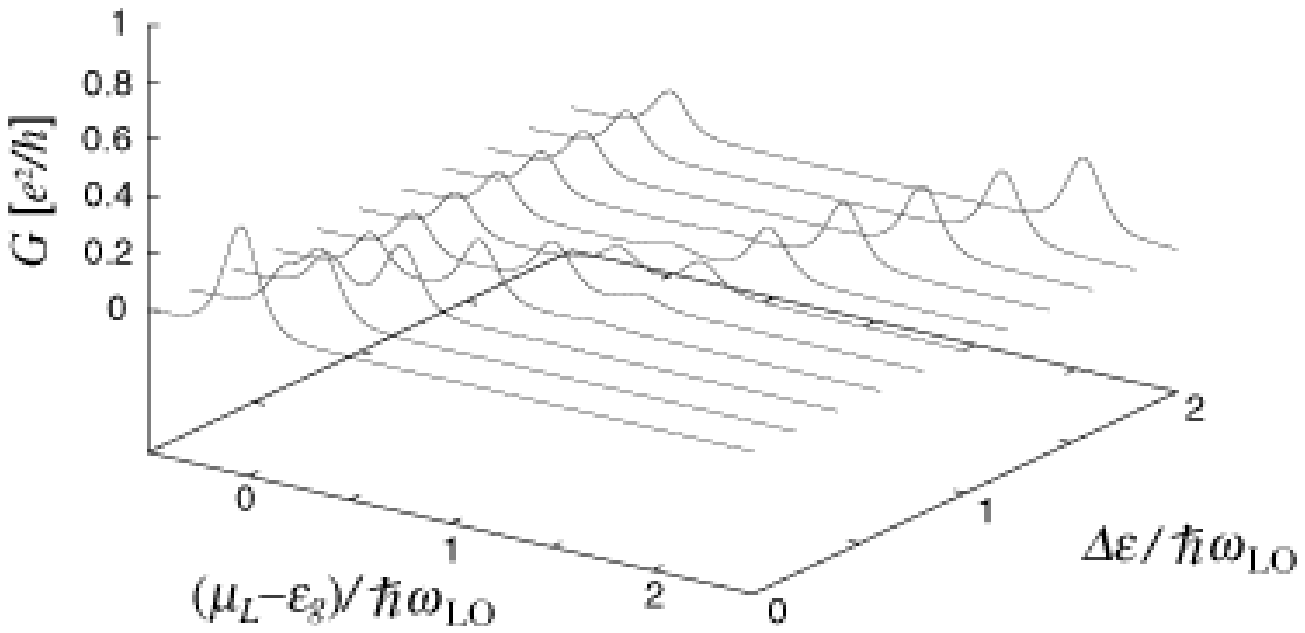}
\end{center}
\caption{The differential conductance $G$ through the quantum dot at
$k_{\rm B}T=0.05\hbar\omega_{\rm LO}$ ($T \approx 18$K),
as a function of the Fermi level in lead $\LL$, $\mu_{\LL}$ ($\mu_{\RL}$
is fixed). The level spacing $\Delta \ene=\varepsilon_e-\varepsilon_g$
in the dot is gradually changed. The level broadenings are the same as
in Fig.\ 3 ($\Delta^g_{\LL}=\Delta^g_{\RL}=0.005\hbar\omega_{\rm LO}$,
$\Delta^e_{\LL}=\Delta^e_{\RL}=0.01\hbar\omega_{\rm LO}$).
}
\label{fig5}
\end{figure}

\vspace{3cm}

\noindent
{\large Tasai and Eto (Fig.\ 5)}


\begin{thebibliography}{99}
\bibitem{Arakawa}
Y.\ Arakawa and H.\ Sakaki: Appl.\ Phys.\ Lett.\ {\bf 40} (1982) 939.
\bibitem{Bastard0}
U.\ Bockelmann and G.\ Bastard: Phys.\ Rev.\ B {\bf 42}
(1990) 8947.
\bibitem{Benisty}
H. Benisty, C. M. Sotomayor-Torres and C. Weisbuch:
Phys.\ Rev.\ B {\bf 44} (1991) R10945.
\bibitem{relax0}
U.\ Bockelmann and T.\ Egeler: Phys.\ Rev.\ B {\bf 46} (1992) 15574.
\bibitem{relax1}
T.\ Inoshita and H.\ Sakaki: Phys.\ Rev.\ B {\bf 46} (1992) 7260.
\bibitem{relax2}
T.\ Inoshita and H.\ Sakaki: Physica B {\bf 227} (1996) 373.
\bibitem{relax3}
X.\ Q.\ Li, H.\ Nakayama and Y.\ Arakawa: Phys.\ Rev.\ B {\bf 59} (1999) 5069.
\bibitem{relax4}
A.\ V.\ Uskov, J.\ McInerney, F.\ Adler, H.\ Schweizer and M.\ H.\ Pilkuhn:
Appl.\ Phys.\ Lett.\ {\bf 72} (1998) 58.
\bibitem{relax5}
O.\ Verzelen, R.\ Ferreira and G.\ Bastard:
Phys.\ Rev.\ B {\bf 62} (2000) R4809.
\bibitem{Inoshita}
T.\ Inoshita and H.\ Sakaki: Phys.\ Rev.\ B {\bf 56} (1997) R4355.
\bibitem{Bastard1}
S.\ Hameau, Y.\ Guldner, O.\ Verzelen, R.\ Ferreira, G.\ Bastard,
J.\ Zeman, A.\ Lemaitre and J.\ M.\ Gerard:
Phys.\ Rev.\ Lett.\ {\bf 83} (1999) 4152.
\bibitem{Bastard2}
O.\ Verzelen, S.\ Hameau, Y.\ Guldner, J.\ M.\ Gerard, R.\ Ferreira
and G.\ Bastard: Jpn.\ J.\ Appl.\ Phys.\ {\bf 40} (2001) 1941.
\bibitem{QD1}
M.\ Narihiro, G.\ Yusa, Y.\ Nakamura, T.\ Noda and H.\ Sakaki:
Appl.\ Phys.\ Lett.\ {\bf 70} (1997) 105.
\bibitem{QD2}
D.\ G.\ Austing, S.\ Tarucha, P.\ C.\ Main, M.\ Henini, S.\ T.\ Stoddart
and L.\ Eaves: Appl.\ Phys.\ Lett.\ {\bf 75} (1999) 671.
\bibitem{Kawamura}
K.\ Kawamura and T.\ Aono: Jpn.\ J.\ Appl.\ Phys.\ {\bf 36} (1997) 3951.
\bibitem{Fano}
U.\ Fano: Phys.\ Rev.\ {\bf 124} (1961) 1866.
\bibitem{com1}
We assume two channels in the leads; one channel couples to level $g$
in the dot only and the other couples to level $e$.
This assumption avoids the mixing between levels $g$ and $e$ through
the tunnel couplings to the leads.
\bibitem{Mahan}
G.\ D.\ Mahan: {\it Many-Particle Physics}, second edition (Plenum Press,
New York, 1990).
\bibitem{com2}
In section 4, we calculate the Green's function for electrons.
The states for the phonons are summed up in the self-energy.
This is in contrast to the self-energies in Eqs.\ (\ref{eq:selfE1}),
(\ref{eq:selfE2}) in section 3 where we consider the Green's function
for the complex system of electrons and phonons.
\bibitem{com4}
Equation (\ref{eq:conductance}) in section 3 implies the suppression
of the peak height. In Fig.\ 2(b), however, this suppression of $G$ is
not observed since Im$\Sigma_{(e, 0_{\rm ph})}=0$ at the peak positions.
\end{thebibliography}
\end{document}